\documentclass[12pt,amsmath,amssymb,aps
notitlepage,
 reprint,
 superscriptaddress,
 showkeys
]{revtex4-2}
\usepackage{graphicx}% Include figure files
\usepackage{hyperref}
\usepackage{epsfig}
\usepackage{xcolor} %example: {\color{red} text}

\newcommand{\bra}[1]{\langle #1 \vert}
\newcommand{\ket}[1]{\vert #1 \rangle}

\begin{document}

\preprint{APS/123-QED}

\title{Resonant cavity-QED with chiral flat bands} 

\author{E. M. Broni}
\affiliation{%
 Instituto de F\'{i}sica, Universidade Federal de Alagoas, 57072-900 Macei\'{o}, AL, Brazil
}%

\author{A. M. C. Souza}
\affiliation{%
 Departamento de F\'{i}sica, Universidade Federal de Sergipe, 49100-000 S\~{a}o Crist\'{o}v\~{a}o, SE, Brazil
}%

\author{M. L. Lyra}
\affiliation{%
 Instituto de F\'{i}sica, Universidade Federal de Alagoas, 57072-900 Macei\'{o}, AL, Brazil
}%

\author{F. A. B. F. de Moura}
\affiliation{%
 Instituto de F\'{i}sica, Universidade Federal de Alagoas, 57072-900 Macei\'{o}, AL, Brazil
}%

\author{G. M. A. Almeida}
\email{gmaalmeida@fis.ufal.br}
\affiliation{%
 Instituto de F\'{i}sica, Universidade Federal de Alagoas, 57072-900 Macei\'{o}, AL, Brazil
}%

\begin{abstract}  
 Flat bands exhibit high degeneracy and intrinsic localization, offering a promising platform for enhanced light-matter interactions. Here, we investigate the resonant interaction between a two-level emitter and a chiral flat band hosted by a photonic lattice. 
In the weak coupling regime, the emitter undergoes Rabi oscillations with a lifted photonic mode whose spatial structure reflects the nature of compact localized states and the onset of Anderson localization. 
We show that weak hopping disorder induces a delocalization of the lifted mode
whereas the effective emitter-field coupling strength, and the associated mode volume experienced by the emitter, remains protected against structural fluctuations.
We illustrate our approach using selected flat band lattices. Our findings provide a route to flat band state preparation via quench dynamics and robust cavity-QED control.
\end{abstract}

\maketitle 

\section{Introduction}

Certain lattices are known to host flat bands, that is, dispersionless bands  
extending over the Brilloin zone resulting in zero group velocity and diverging DOS \cite{leykam18-1,leykam18-2,vicencio21}.
This high level of degeneracy, combined with lattice symmetries, allows for a description of flat-band modes in terms of so-called compact localized states (CLSs), 
which have spatial support limited to a finite number $U\geq 1$ of adjacent unit cells \cite{maimaiti17,maimaiti19}. 
%Indeed, they are classified according to the number of cells $U\geq 1$ they span. 
Since any superposition of states that compose a degenerate set yields a valid eigenstate, CLSs are often
taken as those states with the lowest $U$ that decouple from the rest of the lattice.
Those belonging to class $U=1$ are trivial as they do not overlap and thus form an
orthogonal set. 
%
% Talk about various applications with flat bands and then mention experimental realizations (photonics being the most important one). 
As such, flat bands exhibit insulator characteristics even in the complete absence of disorder -- unlike Anderson localization \cite{anderson58} -- and can be embedded across a dispersive band or separated by a gap \cite{flach14,calugaru22}, yielding unusual localization properties \cite{leykam17}. 
These and other characteristics confer upon flat-band lattices exotic strongly-correlated phenomena \cite{bodyfelt14,derzhko15,peotta15,taie15,li22,hu23,martinez23,chen25}.
%particularly so when the degeneracy is affected by perturbations \cite{khomeriki16}.
%furthermore...
%especially near the point of degeneracy lifting
 
While flat bands can naturally emerge in various materials \cite{derzhko15,elias21,hu23}, rapid progress has been achieved with artificial flat bands occurring in quasi-1D and 2D lattices across plataforms such as electrical lattices \cite{mayoral24}, ultracold atoms \cite{jo12,taie15,li22}, circuit QED \cite{martinez23}, Rydberg lattices \cite{chen25}, and photonics \cite{leykam18-1,vicencio21}. 
%% microwave resonators: Realization of all-band-flat photonic lattices
%
State-of-the-art technology in photonic lattices enables the implementation of tight-binding models with high degree of tunability and local addressing \cite{danieli24}. 
Arrays of laser-written coupled waveguides, for instance, have been successfully employed to realize a variety of lattice geometries containing tens of sites, including, to name a few, Lieb \cite{vicencio15,mukherjee15}, diamond \cite{mukherjee15-2}, and stub \cite{real17,aravena22} lattices.
%Realizations of flat bands have also been reported in photonic crystals \cite{}, thereby raising possibilities in the integration of photonic flatbands with paving the way for better suited interfacing between
%

Inspired by recent developments in photonic crystals \cite{schulz17,nguyen22,le24,yang23,wang25}, there has been growing interest in the role of flat bands in enhancing light–matter interactions \cite{bienias22,yang23, bernardis23,dibenedetto25,wang25}. The associated large DOS with 
zero bandwidth make flat bands functionally analogous to high-Q cavities with effectively small mode volumes.
Also, because the CLSs spread across the lattice, flat band modes offer large tolerance to the atom's position.
Within this context, the study in Ref. \cite{dibenedetto25} characterizes atom-photon bound states arising when an emitter is dispersively coupled to a flat-band. As the detuning decreases, the localization length saturates to a level depending on the overlap between CLSs belonging to classes $U>1$. This behavior contrasts with the case of an emitter approaching the edge of a dispersive band, where a delocalized photonic wavefunction typically emerges \cite{sundaresan19}.
Thus, the intrinsic localization mechanism of flat bands can manifest in cavity-QED platforms, enabling access to challenging regimes of light-matter interaction.
%which indeed makes them a promising platform for realizing cavity-QED phenomena.
For example, a recent experimental realization of a quantum dot interacting with 
a moir\'e flat band cavity has demonstrated strong enhancement and inhibition of the Purcell effect \cite{wang25}.
%Another remarkable trait of flat bands 

Motivated by the rapid progress in the field, here we investigate the resonant interaction between a two-level emitter a chiral flat band \cite{ramachandran17} supported by a photonic lattice.
In the weak coupling regime, 
we derive an effective interaction between the emitter and a well-defined mode that lifts from the flat band, whose spatial profile depends on the subtle tradeoff between compact (intrinsic) and Anderson localization. 
This competition produces a characteristic disorder-induced delocalization of the lifted mode, which can be probed through the Rabi dynamics triggered by the atomic emission.
%
%The corresponding Rabi frequency is the square root of the population of the site to which the emitter is coupled in the flat band.
It occurs at a frequency proportional to the square root of the total flat-band weight on the site to which the emitter is coupled.
Furthermore, we show that the effective mode volume experienced by the emitter remains remarkably stable against hopping disorder, despite the substantial reshaping of the lifted photonic mode.

We illustrate these results using a few paradigmatic flat band models, highlighting features such as symmetry-protected CLSs and disorder-induced delocalization.
%
%quantum probing \cite{}, in which a small system is used to control or extract information from a larger, more complex one.
%
Furthermore, implied in our work is the perspective of preparing flat band states via quench dynamics, rather than through their direct excitation, which can be problematic when the CLSs are non-orthogonal \cite{mayoral24}. As such, we aim to provide a simple model that captures the essential mechanism and does not disturb the flat band even in the presence of disorder.

\section{Model}

Let us begin by considering a two-level quantum emitter, with ground state $\ket{g}$ and excited state $\ket{e}$ separated by frequency $\omega_e$, interacting with a photonic lattice. 
The latter is expressed in the form of coupled 
lossless cavities
with overlapping spatial modes, as given by the tight-binding Hamiltonian ($\hbar=1$)
\begin{equation}
\hat{H}_{\mathrm{field}} =\sum_{ x\neq x' }J_{x,x'} (\hat{a}_{x'}^{\dagger}\hat{a}_{x}+\mathrm{h.c.}),
\end{equation}
where $J_{x,x'}$ is the hopping strength (expressed in units of $J\equiv1$) and $\hat{a}_x$ $(\hat{a}_x^{\dagger})$ is the bosonic annihilation (creation) operator acting at $x$th cavity. 
As we will focus on flat bands protected by chiral (or sublattice) symmetry \cite{ramachandran17}, all the cavity frequencies are the same and set to zero for simplicity.
Assuming that the emitter couples directly to the cavity located at $x_0$, undergoing Jaynes-Cummings interaction in the rotating wave approximation, the full Hamiltonian reads
\begin{equation} \label{full}
\hat{H} =\omega_e \ket{e}\bra{e}+\hat{H}_{\mathrm{field}}+g_0(\hat{\sigma}_{+}\hat{a}_{x_0}+\mathrm{h.c.}),
\end{equation}
where $g_0$ is the atom-cavity coupling strength and $\sigma_{+}=\ket{e}\bra{g}$ is the atomic raising operator. 
%

%%%%%%%%%%%%%%%%%%
The system Hamiltonian preserves the total number of excitations. 
By initializing the system as $\ket{\Psi(0)}=\ket{e}\ket{\mathrm{vac}}$, with $\ket{\mathrm{vac}}$ being the field vacuum state, we get $\ket{\Psi (t)} = e^{-i \hat{H}t}\ket{\Psi(0)}= f_e(t)\ket{e}\ket{\mathrm{vac}}+ \sum_x f_x(t)\ket{g}\ket{x},
$
where $f_e(t)$ ($f_x(t)$) is the emitter (field) amplitude and $\ket{x}=\hat{a}_x^{\dagger}\ket{\mathrm{vac}}$ are single-photon Fock states. As such, our analysis is restricted to the single-excitation sector. From now on, we write $\ket{e}\ket{\mathrm{vac}}\rightarrow \ket{e}$ and $\ket{g}\ket{x} \rightarrow \ket{x}$ for short.   
%%%%%%%%%%%%%%%%%%%%%

To see how the emitter interacts with the lattice modes, let us express the field Hamiltonian in terms of its 
eigenstates as $\hat{H}_{\mathrm{field}} = \sum_{\mu,k}\omega_{\mu}(k)\ket{\psi_{\mu,k}}\bra{\psi_{\mu,k}}$, where the eigenvalues $\omega_{\mu}(k)$ form the band structure, with $\mu$ being the band index and $k$ the wavenumber defined in the first Brillouin zone. In this picture, a quick inspection shows that the emitter couples
to each field mode at a rate $g_{\mu,k}  \equiv g_0\langle x_0 \vert \psi_{\mu,k} \rangle$, which is assumed real \cite{lorenzo17, monteiro25}. Now, we consider that one of the bands, say $\mu'$, is completely flat, i.e., $\omega_{\mu'}(k) = \omega_{\mathrm{FB}} = \mathrm{constant}$, and 
detuned from the closest dispersive band(s) by $\Delta$. Fixing $\omega_e=\omega_{\mathrm{FB}}$ and taking $g_0\ll \Delta$ such that the emitter is finely tuned to the flat band but has negligible coupling with the remaining modes, we end up with the effective Hamiltonian to first order in $g_0$ (offset by $\omega_\mathrm{FB}$): 
\begin{equation} \label{Heff}
\hat{H}_{\mathrm{eff}}=\sum_k g_k(\ket{e}\bra{\psi_{k}}+\mathrm{h.c.}),
\end{equation}
where $\ket{\psi_{k}}$ are the corresponding flat-band modes. From now on, we are locked into this interaction regime and will omit the band index $\mu$ for brevity. 

% Disorder and symmetry
Remarkably, the above description holds even in the presence of off-diagonal disorder -- say, with $J_{x,x'}$ drawn from a uniform random distribution of width $W$ --
%say, with $J_{x,x'}\propto(1+\delta)J$, with $\delta \in [-W/2,W/2]$ randomly picked from a uniform distribution of width $W$ -- 
provided that the lattice is bipartite and the flat band occurs at $\omega_\mathrm{FB}=0$ \cite{ramachandran17}.
According to Lieb’s theorem \cite{lieb89}, a zero-energy flat band is guaranteed in lattices with chiral symmetry and an odd number of sites per unit cell.
%
%We will thus consider $J_{x,x'}\rightarrow J_{x,x'}(1+\delta)$ independently drawn from a random distribution $[-W,W]$, $W$ being the disorder strength.
%
Another important theorem \cite{sutherland86,inui94}
%particularly useful for engineering chiral flat bands \cite{ramachandran17}, 
states that \emph{at least} $M-m$ zero-energy modes are always present in bipartite lattices, where $M$ ($m$) is the number of sites belonging to the majority (minority) sublattice. These modes have no support on the minority sublattice. 

\begin{figure} 
    \includegraphics[width=0.45\textwidth]{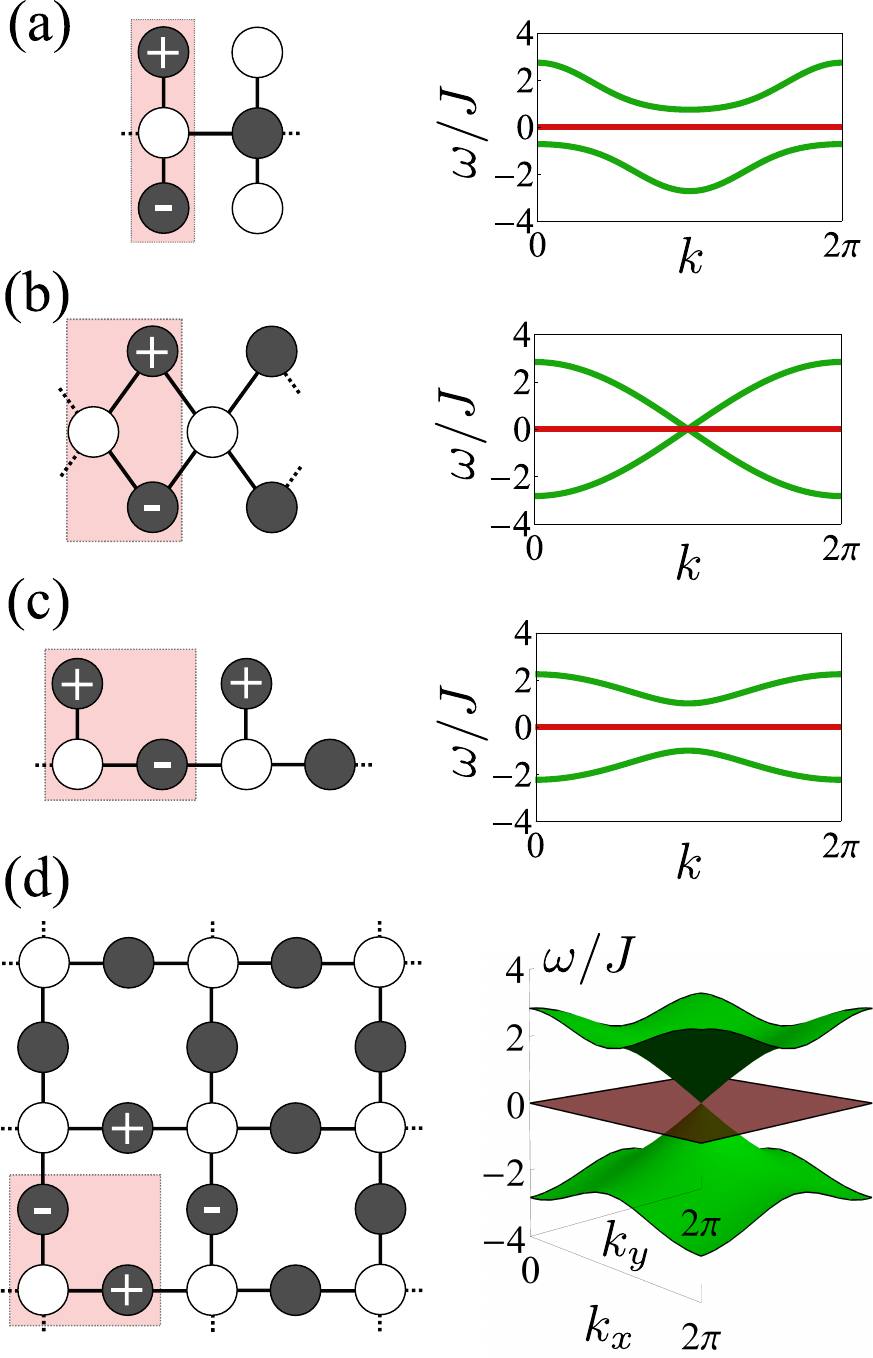}
    \caption{Flat-band lattices and their corresponding frequency spectra.
    Filled and empty sites highlight the bipartite structure of the lattices. All couplings are uniform, $J_{x,x'}=J$, and the signs $\pm$ stand for the CLS amplitudes.
    In this work, we showcase the cavity-QED physics in the (a) double-comb ($U=1$) \cite{dibenedetto25}, (b) diamond ($U=1$) \cite{mukherjee15-2}, (c) stub ($U=2$) \cite{real17}, and (d) 2D Lieb ($U=3$) \cite{nita13} lattices. In the presence of coupling disorder, the flat band is maintained at zero energy but the CLS class may be modified.  
    } 
\label{fig0}
\end{figure}

Instances of bipartite flat-band lattices and their corresponding energy spectrum are displayed in Fig. \ref{fig0}. Those will be discussed in detail to showcase our main results shortly. The CLS structure in the homogeneous case is depicted for each lattice. Even though in some cases the flat band touches the dispersive bands, as in the diamond and 2D Lieb lattices  [see Figs. \ref{fig0}(b) and \ref{fig0}(d)], we will be dealing with finite-size systems, securing the existence of a gap $\Delta$. Nevertheless, the dispersive part of the spectrum may contribute with additional degeneracies, which can be removed by the disorder or carefully setting the parity of the number of unit cells so as to forbid certain values of $k$ \cite{ramachandran17,nita13}. Otherwise, access to the pure flat band is compromised within the proposed cavity-QED approach, as the minority sublattice would participate in the atom-field exchange described next.

%and it also restricts $x_0$ to one of the cavities in the majority set.
%
%Case examples are given shortly. 

\section{Effective Jaynes-Cummings dynamics}

For arbitrary couplings $g_k$, we can diagonalize Eq. (\ref{Heff}) to obtain the pair of hybrid light-matter eigenstates
\begin{equation} \label{psi_lambda}
\ket{\phi_\pm}=\frac{1}{\sqrt{2}}\left(\ket{e}\pm\frac{1}{\lambda}\sum_kg_k\ket{\psi_k}\right),
\end{equation}
with the Rabi splitting (effective coupling strength) given by $\pm\lambda=\pm\sqrt{\sum_k g_k^2}$. Assuming that the lattice has $N$ unit cells (and therefore $N$ flat-band states), the remaining $N-1$ zero-energy eigenstates do not have amplitude on $\ket{e}$ and therefore do not participate in the emission dynamics. This can be seen by realizing that the effective Hamiltonian itself describes a bipartite star network, with $\ket{e}$ alone representing the minority sublattice. 
%The symmetry also establishes that 1/2 of the probability distribution 
%

According to Eq. (\ref{psi_lambda}), the unitary time evolution of 
$\ket{\Psi(0)} = \ket{e}$ yields Rabi oscillations with the lifted mode 
\begin{equation}
\ket{\widetilde{\psi}}\equiv \lambda^{-1}\sum_k g_k \ket{\psi_k}=\frac{\mathcal{P}\ket{x_0}}{\sqrt{\langle x_0 |\mathcal{P}|x_0 \rangle}},
\end{equation}
with $\mathcal{P}=\sum_k \ket{\psi_k}\bra{\psi_k}$ being the projector onto the flat-band subspace. 
The same form for the lifted mode was obtained by Di Benedetto \textit{et al.} in Ref. \cite{dibenedetto25}, where the authors examined the localization properties of atom–field bound states with the emitter coupled off-resonantly to the flat band.

Here, we focus on the resonant limit, in which the atom fully releases its energy at odd multiples of $\tau=\pi/(2\lambda)$, i.e., $f_e(\tau)=0$. 
The frequency $\lambda$ depends on the total weight of the $x_0$ amplitudes within the flat band. We will see later that this quantity is quite robust against hopping disorder. Note that the probability $|\langle x_0\ket{\widetilde{\psi}}|^2=(\lambda/g_0)^2$. 

In general, the formalism above applies to any degenerate level separated by a finite gap 
$\Delta \gg g_0$, where the momentum $k$ is replaced by some other label in Eq. (\ref{Heff}).
We note that while the disorder is expected to broaden the dispersive bands, it cannot close the (pseudo)gap \cite{ramachandran17}.  
Given a real Hamiltonian with sublattice symmetry,
energy levels repel around zero energy, ensuring no crossings. 
%In fact, a topological transition would take place if the levels crossed zero energy. 
%
We checked numerically that 
setting $g_0=10^{-3}J$ ensures $\Delta \gg g_0$ by at least two orders of magnitude. 
%Here, we fix $g_0=10^{-3}J$, which fulfills such condition even in the presence of disorder, where a pseudogap surges instead \cite{ramachandran17}.

\section{Spatial properties of the lifted mode $\ket{\widetilde{\psi}}$}

Now, the question that follows is: which linear combination of flat band modes, $\ket{\widetilde{\psi}}$, does
the emitter ``choose'' to interact with? 
If $\ket{\widetilde{\psi}}$ corresponded to a single, non-degenerate mode, the answer would be straightforward. However, there are no obvious constraints on the spatial structure of the lifted mode unless $U=1$ CLSs are supported by the flat band. In this case, all $g_k$ vanish except for the one that matches with the CLS
defined at the cell containing the emitter. Hence, $\ket{\widetilde{\psi}}$ assumes the form of the CLS \cite{dibenedetto25}. 

Below, we analyze the spatial profile of the lattices displayed in Fig. \ref{fig0} and the role played by the corresponding $U$ class, flat-band support at $x_0$, and Anderson localization. Unless stated otherwise, hopping disorder is set as $J_{x,x'}\rightarrow (1+\delta_{x,x'})J$, where each $\delta_{x,x'}\in[-W/2,W/2]$ is drawn independently from a uniform random distribution of width $W$.
We perform such analysis by exploiting the resonant cavity-QED dynamics triggered by the emitter. The connection between the full Hamiltonian [Eq. (\ref{full})] and the effective interaction described by Eqs. (\ref{Heff}) and (\ref{psi_lambda}) is validated by $\langle x\ket{\widetilde{\psi}} \equiv f_x(\tau)$. 

To quantify the degree of localization, we use the participation ratio defined as $\xi=(\sum_{x}|\langle x \vert \widetilde{\psi} \rangle |^4)^{-1}$, which
ranges from $O(N)$ for delocalized states to $O(1)$ for strongly localized ones. 

% We can immediately argue that if the lattice supports $U=1$ CLSs, then all $g_k$ vanish except the one, $g_{k_0}$, that corresponds to the unit cell containing the emitter at $x_0$. Hence, $\ket{\widetilde{\psi}}$ becomes the CLS $\ket{\psi_{n_0}}$ at that cell.
%

\subsection{Double-comb lattice}

\begin{figure} 
    \includegraphics[width=0.47\textwidth]{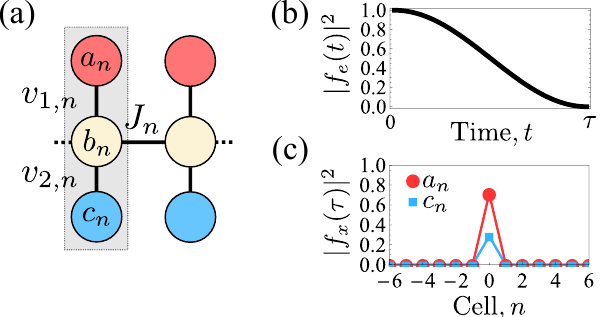}
    \caption{(a) Double-comb lattice with arbitrary hopping strengths, whose flat band hosts
CLSs of the $U=1$ class. The unit cell is indicated by the shaded box. 
(b) Emitter probability amplitude $|f_e(t)|^2$ versus time. The curve is numerically obtained for 
a single realization of hopping disorder with $W=1$. The lattice consists of $N=20$ cells with periodic boundary conditions.
(c) Photonic probability amplitude $|f_x(\tau)|^2\equiv |\langle x \vert \widetilde{\psi} \rangle |^2$ evaluated at time $\tau = \pi/(2\lambda)$. Only $a-$ and $c-$sites are shown as $b-$sites do not contribute to the flat band. Lines are for guiding the eye. 
The CLS is manifested either for 
$x_0=a_0$ or $x_0=c_0$, with distinct Rabi frequencies $\lambda$ (see text).} 
\label{fig1}
\end{figure}

We begin by considering the disordered double-comb lattice displayed in Fig. \ref{fig1}(a). This geometry is interesting because in addition to supporting a symmetry-protected flat band, the CLSs remain in the $U=1$ class for any strength of the hopping disorder. 

Defining $r=v_{2,n}/v_{1,n}$, each cell $n$ supports a CLS of the form 
\begin{equation}\label{dc}
\ket{\psi_n} = ( r\ket{a_n}-\ket{c_n})/\sqrt{1+r^2}.
\end{equation}
If the emitter couples to an $a-$site at cell $n=0$, namely  $x_0=a_0$, then the 
coupling frequency reads $\lambda = g_0r/\sqrt{1+r^2}$, such that the CLS of the corresponding cell is dynamically obtained at time $\tau\propto\lambda^{-1}$, following complete emission of the atom. 
This double-comb lattice thus acts analogously to a chain of uncoupled cavities, each cavity mode being represented by a $\ket{\psi_n}$.
In Figs. \ref{fig1}(b) and \ref{fig1}(c) we display the
evolution of the emitter amplitude $|f_e(t)|^2$
and wavefunction population $|f_x(\tau)|^2$ at time $t=\tau$, respectively.
%Data are obtained from a single realization of random hoppings $(1+\delta)J$, with independent $\delta \in [-W/2,W/2]$.
%

Note that unlike conventional chiral flat bands where zero-energy modes reside exclusively on the majority sublattice according to the theorems in Refs. \cite{sutherland86,inui94}, in the double-comb lattice the CLSs alternate between the two sublattices. The CLS supported on cell $n$ resides on the
$a-$ and $c-$sites belonging to
one sublattice, while that on cell $n+1$ resides on the other sublattice [compare Figs. \ref{fig0}(a) and \ref{fig1}(a)]. 
This behavior reflects the fact that the translational symmetry of the lattice and the sublattice labeling are not commensurate. In this situation the band structure features a glide reflection symmetry, which relates an eigenstate 
at $\omega(k)$ with one at $-\omega(k+\pi)$, rather than $-\omega(k)$ [see right panel of Fig. \ref{fig0}(a)]. 
Accordingly, the double-comb lattice is formally classified as a class-II sublattice-symmetric lattice \cite{xiao24}, rather than a chiral lattice. The equivalency between these two symmetries holds only when the periodicity of the primitive unit cell coincides with that of the sublattice labeling \cite{xiao24}.
%%%%%
As a consequence, the zero-energy degeneracy is not due to a global imbalance between majority and minority sublattices (their numbers are equal), but instead originates from a local interference condition satisfied independently in each unit cell, as expressed in Eq. (\ref{dc}). Each cell therefore contributes one compact localized state to the flat band even in the presence of disorder, and Anderson localization is not manifested.  

\begin{figure} 
    \includegraphics[width=0.47\textwidth]{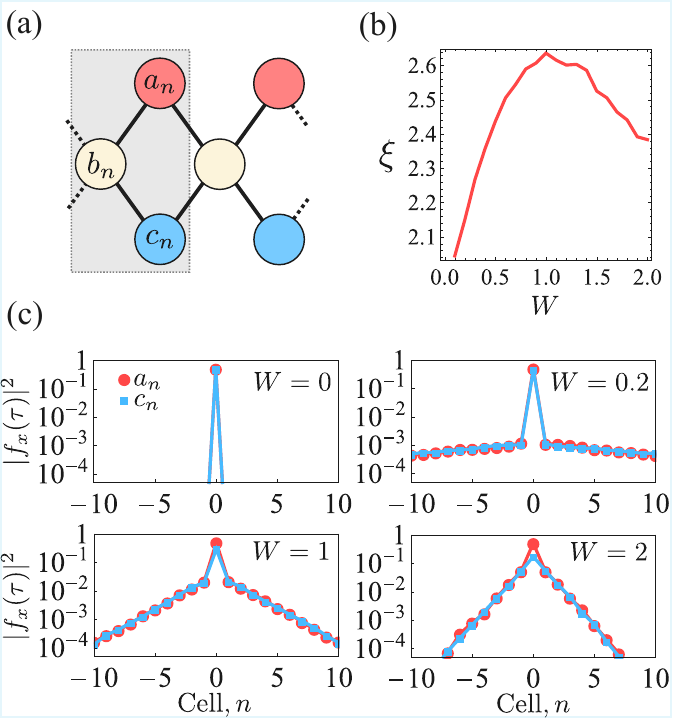}
    \caption{(a) Schematic of the diamond chain. $b-$sites form the minority sublattice and do not contribute to the flat band. 
    (b) Participation ratio $\xi$ versus disorder strength $W$. (c) Photonic population $|f_x(\tau)|^2$ for selected values of $W$. Data are averaged over $10^4$ disorder samples for $N=31$ cells. 
    } 
\label{fig2}
\end{figure}

\subsection{Diamond lattice}

%\begin{equation}
%\ket{\psi_n} = \left( \ket{a_n}-\frac{v_{1,n}}{v_{2,n}}\ket{c_n}+\frac{J_{2,n+1}\epsilon_{1,n}}{J_{1,n+1}\epsilon_{2,n}}\ket{a_{n+1}}-\frac{\epsilon_{1,n}}{\epsilon_{2,n}}\ket{c_{n+1}} \right)
%\end{equation}
% diamond lattice
Next, we examine the diamond (or rhombus) chain \cite{vidal00, doucot02,mukherjee15-2} shown in Fig. \ref{fig2}(a). 
In the homogeneous case, the CLSs belong to the $U=1$ class (for real hopping parameters \cite{khomeriki16}) and read
\begin{equation}\label{clsdiamond}
\ket{\psi_n}=(\ket{a_n}-\ket{c_n})/\sqrt{2}.
\end{equation}
For disordered hoppings, they become $U=2$ as it can be inferred by taking an arbitrary superposition of sites $a_n,a_{n+1},c_{n},c_{n+1}$ in order to obtain a set of 3 equations as 
the 4 unknown components meet at $b_n, b_{n+1},b_{n+2}$ ($b-$sites do not have components in the chiral flat band as they belong to the minority sublattice). The normalization condition eliminates one of these unknowns and the CLS can be defined. The existence of a CLS form is guaranteed for chiral flat-band lattices featuring nearest-neighbor hopping \cite{ramachandran17}.

%$\hat{H}_{\mathrm{field}}$.

CLSs of classes $U>1$ constitute a non-orthogonal set. Hence, the flat-band modes $\ket{\psi_k}$ are linear combinations of such compact states. The lifted mode $\ket{\widetilde{\psi}}$, still only present on the majority sublattice, will inevitably cover all unit cells and feature an exponential tail.   
Note that even in the absence of disorder, atom-photon bound states are typically exponentially localized around the atom when it is
dispersively coupled to either a standard band \cite{sundaresan19} or a flat band \cite{dibenedetto25}. 
 In the latter, the authors derived universal scaling laws for the increase of the localization length with the overlap $\langle \psi_n | \psi_{n\pm1}\rangle = \alpha$ between the CLSs.

In the diamond chain, the hopping disorder acts by shuffling the overlap among distinct cells $\langle \psi_n | \psi_{n+1}\rangle = \alpha_n$. 
To see how this mechanism promotes Anderson localization in the flat band, 
Fig. \ref{fig2}(b) shows the participation ratio $\xi$ versus the disorder width $W$.
We identify a disorder-induced delocalization for weak disorder. This kind of behavior is common in flat-bands systems and can be associated to a crossover from 
intrinsic compact localization into the Anderson regime \cite{li22,wang22,goda06, almeida23flat}.
We remark that, in our coupled-cavity setting, this feature appear in the lifted mode $\ket{\widetilde{\psi}}$ (as a consequence of the bare flat band) and is triggered by the atomic emission process. This makes cavity-QED systems promising platforms to harness and probe flat-band properties.

%
%
%This anomalous localization behavior is also reported in Ref. \cite{almeida23flat}, where the diamond chain is employed as a quantum communication channel. 
In Fig. \ref{fig2}(c), the spatial shape of $|f_x(\tau)|^2$ is displayed for selected values of $W$. It is shown in semi-log scale to highlight the exponential tail of the wavepacket. 
The 
residual contribution of the $U=1$ CLS form [Eq. (\ref{clsdiamond})] is seen for all values of $W$. 
Yet, the localization of the flat-band is weakened as soon as $W\neq 0$ because the non-orthogonality of the $U=2$ CLSs enlarges the tail of the wavefunction.  
Then, at stronger disorder Anderson localization dominates, as noted by the decrease of $\xi$.

Given the periodic boundary conditions to the chain, $N$ must be odd in the clean case. Otherwise, a pair of extra modes from the dispersive bands would join the zero-energy level. While this does not invalidate the weak-coupling limit, where the atom still interacts with a degenerate level [as in Eq. (\ref{Heff})], the lifted mode would no longer exhibit the properties of a purely chiral flat band. Our main goal here is to study the trade-off between compact localization and Anderson localization within the flat band, without interference from accidental degeneracies.
%We also note that the exponential localization runs over both $a-$ and $c-$sites separately, despite the 
%We also remark that as closed boundary conditions are being considered, the number of cells $N$ must be odd, otherwise a couple   
%

%

%We now focus on the tradeoff between compact localization and Anderson localization.

%In the case of the diamond chain, we can safely attribute the behavior seen in Figs. \ref{fig2}(b) and \ref{fig2}(c) to the onset of Anderson localization as in the thermodynamic limit $\Delta \rightarrow 0$. This renders the flat band modes less robust against perturbations.
%First, strong disorder has a direct influence on the decrease of the localization length. 
%We remark, however, that our system is finite and therefore Eq. (\ref{Heff}) still hold as long as $g\ll \Delta$.  

\begin{figure} 
    \includegraphics[width=0.47\textwidth]{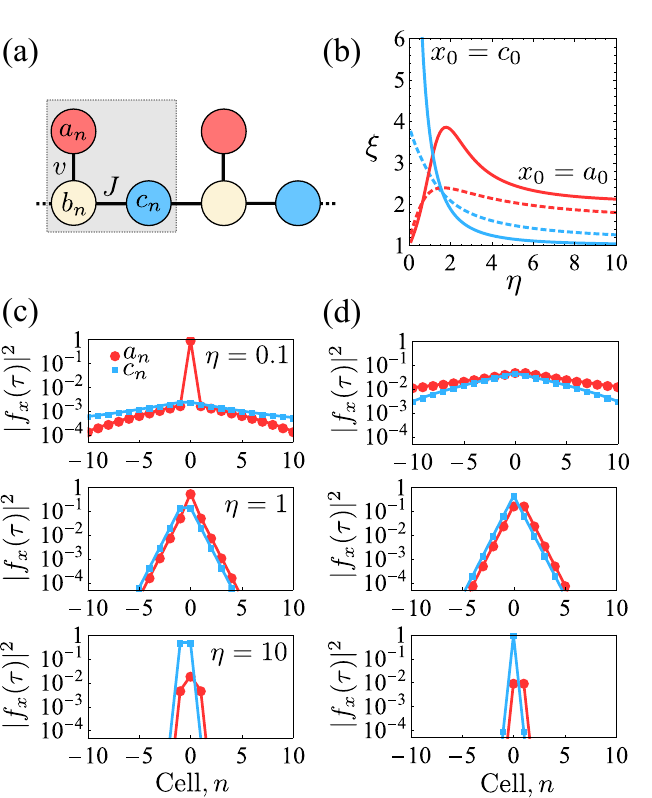}
    \caption{(a) Stub lattice with vertical and horizontal couplings $v$ and $J$, where $\eta = v/J$ controls both the gap between the flat and dispersive bands and the degree of orthogonality of the supporting $U=2$ CLSs. (b) Participation ratio $\xi$ against $\eta$ for both possible locations $x_0$ of the emitter. Solid lines depict the cases without disorder. Dashed lines represent the disordered stub lattice with $v\rightarrow (1+\delta')v$ and $J\rightarrow (1+\delta)J$, where $\delta,\delta' \in [-1,1]$ are random values 
    independently assigned to each coupling throughout the lattice. Results are obtained for $N=31$ cells with periodic boundary conditions and averaged over $10^3$ realizations of the disorder. The spatial profile of the lifted photonic mode is shown for distinct values of $\eta$ in the absence of disorder considering (c) $x_0=a_0$ and (d) $x_0=c_0$.} 
\label{fig3}
\end{figure}

\subsection{Stub lattice}

Next, we explore another $U=2$ flat-band lattice, with a controllable global overlap $\alpha$. Let us analyze the stub lattice \cite{real17,aravena22} depicted in Fig. \ref{fig3}(a). This geometry supports a chiral flat band whose CLSs belong to the $U=2$ class. They are written as 
\begin{equation}\label{clsstub}
\ket{\psi_n}=\frac{\ket{a_n}+\ket{a_{n+1}}-\eta\ket{c_n}}{\sqrt{2+\eta^2}},
\end{equation}
where $\eta=v/J$ is proportional to the gap $\Delta$. Again, $b-$sites span the minority sublattice and do not contribute to the flat band.

The overlap between adjacent CLSs overlap reads $\alpha =(2+\eta^2)^{-1}$, such that orthogonality is reached for $\eta\rightarrow \infty$. In the presence of disorder, it is straightforward to see that the $U=2$ class is preserved, although it differs from Eq. (\ref{clsstub}).
The global parameter $\alpha$ determines the localization length of $\ket{\widetilde{\psi}}$ when $W=0$ \cite{dibenedetto25}. 
Now, the influence of the CLS is non-trivial and depends on whether $x_0$ is an $a-$site or a $c-$ site [see solid curves in Fig. \ref{fig3}(b)]. The corresponding photonic modes are shown in Figs. \ref{fig3}(c) and \ref{fig3}(d) for representative values of $\eta$.
For very small $\eta$, corresponding to a large overlap, we observe strong localization (delocalization) when $x_0=a_0$ ($x_0=c_0$). The former can be understood by noting that 
the flat band description in terms of $U=2$ CLSs is not suitable to capture 
such behavior due to the high overlap between them. 
%Note that $\langle \psi_k \vert \psi_{k'} \rangle = 0$ must be fulfilled in Eq. (\ref{psi_lambda}).
%
In fact, in the limit $\eta \rightarrow 0$, all the $a-$sites are decoupled from the lattice. A proper choice in this case is to write an orthogonal set of (non-compact) states of the form $\ket{\chi_n} \propto \ket{a_n}+\eta \ket{\beta_n}$, where $\ket{\beta_n}$ involves the $c-$sites and remaining $a-$sites and fulfills
$\langle \beta_n \vert \beta_{n'} \rangle=0$. In contrast, when $x_0=c_0$, 
the influence of the Bloch modes becomes evident due to the
proximity of the flat band to the dispersive ones (it actually touches them at $\eta \rightarrow 0$) rendering a large localization length. For the same reason, 
the photonic wavefunction is more sensitive to disorder, as indicated by the dashed curves in Fig. \ref{fig3}(b).

%which is also comparable to the one obtained for the diamond chain under strong $W$ [cf. Fig. \ref{fig2}(c)]. 
For $\eta \sim 1$, both initial conditions yield similar wavefunctions exhibiting exponential localization, despite the absence of disorder. Also, we note that 
a similarity with the diamond chain can be drawn from the   
non-monotonic behavior of $\xi$ versus $\eta$ (in place of $W$) when $x_0=a_0$ [compare Figs. \ref{fig2} and \ref{fig3}]. 
Interestingly, such comparison also suggests that the weak-disorder limit of the diamond lattice is analogous to having a large global $\alpha$.
In the stub lattice,
as the orthogonality of the $U=2$ CLSs 
$\ket{\psi_n}$ becomes dominant (large $\eta$), $\ket{\psi_n}\simeq \ket{c_n}$, leading to $\ket{\widetilde{\psi}}\simeq \ket{c_0}$ ($\ket{\widetilde{\psi}}\simeq \ket{c_{-1}}+\ket{c_0}$) for $x_0=c_0$ ($x_0=a_0$) as visualized in the last panels of Figs. \ref{fig3}(c,d). Only then does the strictly compact form of the flat band modes become dynamically manifested.

\subsection{2D Lieb lattice}

\begin{figure} 
    \includegraphics[width=0.47\textwidth]{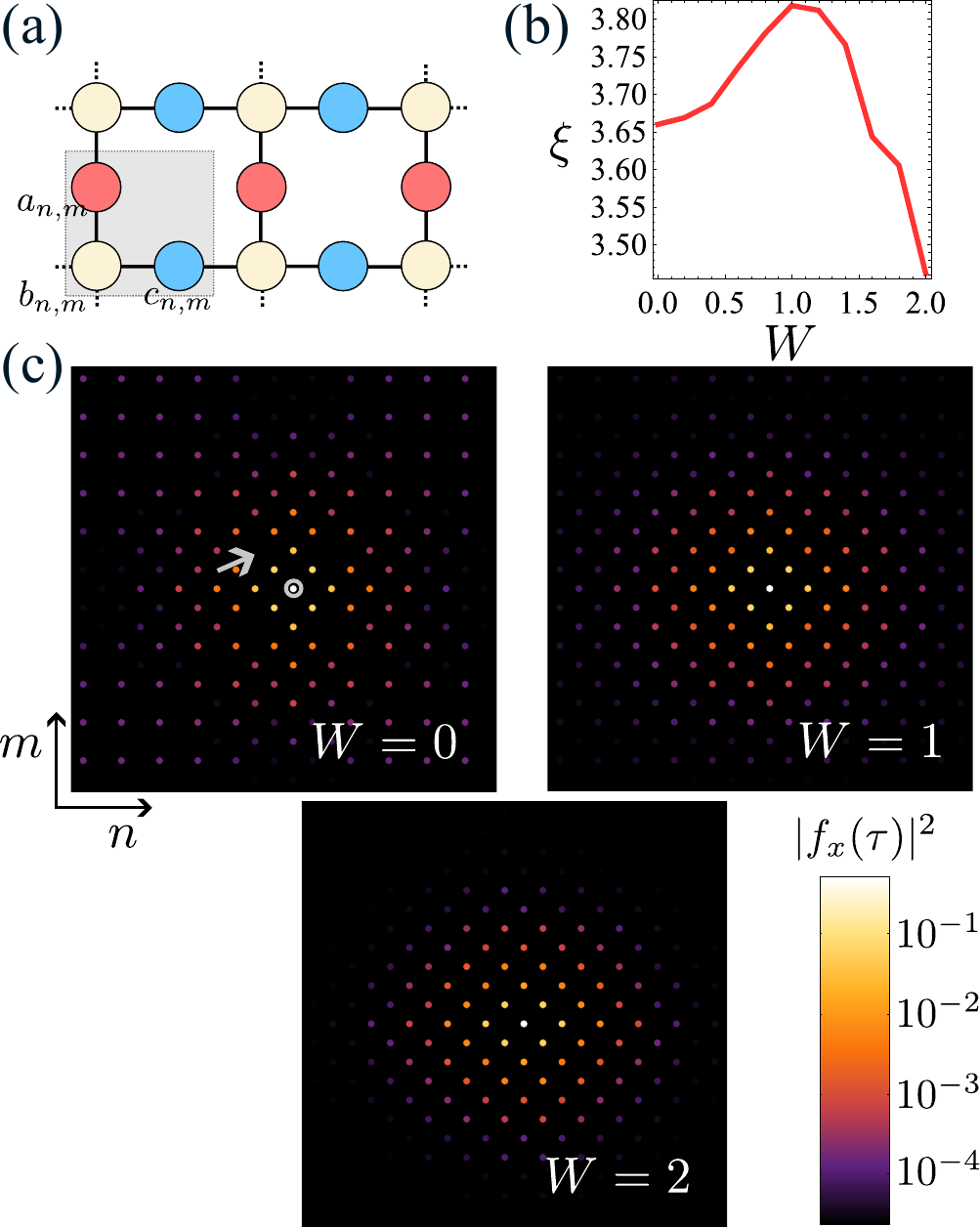}
    \caption{(a) 2D Lieb lattice. Coordinates of the unit cells are $(n,m)$. 
    (b) Participation ratio $\xi$ versus disorder strength $W$, averaged over $10^3$ independent realizations of the disorder. 
    A lattice featuring $N=11\times11$ cells with periodic boundary conditions is considered. (c) Corresponding photonic wavefunction $|f_x(\tau)|^2$ of the lifted mode for distinct values of $W$ (color intensities are shown using a nonlinear asinh scale). The atom is loaded at $x_0$ indicated by the circle. The arrow points to a dark spot (there are many) resulting from symmetry-related destructive interferences in the clean limit.
    } 
\label{fig4}
\end{figure}

To show that our findings also applies to 2D structures, we investigate the Lieb lattice \cite{vicencio15,mukherjee15} displayed in Fig. \ref{fig4}(a). In the clean case, the associated CLSs belong to the $U=3$ class and have the form 
\begin{equation}
\ket{\psi_{n,m}} = (-\ket{a_{n,m}}+\ket{c_{n,m}}-\ket{a_{n+1,m}}+\ket{c_{n,m+1}})/2
\end{equation}
The overlap between two neighboring states (either horizontally or vertically) is then $|\alpha|=1/4$, which is the maximum value allowed for 2D \cite{dibenedetto25}. 

In the disordered case, the CLS class changes to $U=5$. 
Note that if one insists on maintaining the CLS living on a single plaquete, the linear system of equations become overdetermined, yielding correlations between the hopping amplitudes. This is actually worth exploring further on its own in the broader context of localization under correlated disorder \cite{ongoing}. Therefore, if we include an adjacent plaquete (totaling seven sites of the majority sublattice) as shown in Fig. \ref{fig4}(a) then we get six unknowns and six equations (corresponding to the number of $b-$ sites in the region). It is straightforward (though tedious) to derive their analytical form but this is beyond the scope of our work.

Once again, the chiral symmetry-preserving disorder will randomize the overlaps between the CLSs, leading to Anderson localization effects. Figure \ref{fig4}(b) shows the participation ratio versus disorder for a $11\times11$ Lieb lattice with periodic boundary conditions, where a disorder-induced delocalization is obtained, similar to 
that of Fig. \ref{fig2}(b).
This result shows that even when the CLSs maximally overlap in the clean case -- leading to a significantly large localization length \cite{dibenedetto25} -- moderate disorder slightly delocalizes the lifted mode. 
Spatial profiles of the occupation probability
are shown in Fig. \ref{fig4}(c) for selected values of $W$. In the clean case ($W=0$), destructive interferences resulting from the superposition of $\ket{\psi_{n,m}}$ yields dark spots surrounding $x_0$, an instance of which is pointed out by the arrow in Fig. \ref{fig4}(c).
For $W=1$, we see that the disorder turns them into occupied 
sites as the original $U=3$ CLSs are replaced by $U=5$ states with random overlap $\alpha_i$. 
Under strong disorder ($W=2$), 
the mode becomes Anderson localized.

\section{Effective mode volume versus disorder}

\begin{figure} 
    \includegraphics[width=0.47\textwidth]{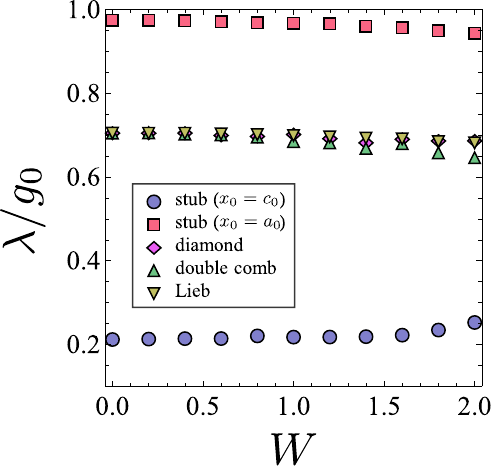}
    \caption{Effective atom-field coupling strength, that is the amplitude of the lifted mode at $x_0$, $\lambda/g_0 =|\langle x_0\ket{\widetilde{\psi}}| $, versus $W$ for the lattices considered in this work. Lattice parameters are the same used in the previous figures. For the stub lattice, we set $\eta=0.1$.
    } 
\label{fig5}
\end{figure}

So far we have explored how the photonic amplitudes are distributed in the flat-band lattices upon
triggering the emission process of an embedded atom. 
One important result is the disorder-induced delocalization
for weak disorder whereas Anderson localization sets in accompanied by the 
increasing randomization of the overlaps $\langle \psi_n | \psi_{n+1}\rangle = \alpha_n$ between CLSs.

Although disorder redistributes the amplitudes throughout the lattice and modifies the CLS structure, we find that the field amplitude at the emitter's location $x_0$ does not respond as much. In Fig. \ref{fig5} we plot the effective atom-field coupling $\lambda =g_0|\langle x_0\ket{\widetilde{\psi}}|$ versus $W$ for the various flat-band lattices.
Under strong disorder, there is a noticeable change for the double-comb and stub lattices
because the increase in $W$ affects the ratio between horizontal and vertical couplings of the unit cell [see Figs. \ref{fig1}(a) and \ref{fig3}(a)].

In cavity-QED, we can associate the atom–field coupling with the inverse square root of an effective optical mode volume $V_{\mathrm{eff}}$. Here, although $g_0$ sets the local coupling with the physical cavity, from the point of view of the emitter, such mode volume is therefore not a geometric property of the cavity but defined by the local amplitude of the particular eigenmode lifted from the flat-band, $\lambda \propto 1/\sqrt{V_{\mathrm{eff}}}$.
Therefore, any increase in the volume would imply that the flat-band is losing support at the emitter location $x_0$. Remarkably, Figure \ref{fig5} shows that 
$\lambda$ can be predicted reliably 
even in devices where fabrication imperfections are unavoidable. A symmetry-protected flat-band lattice acts as a self-stabilizing photonic environment where the destructive-interference constraints protect the local field amplitude seen by the atom.

\section{Conclusions}

%concluding remarks and discussion

We have investigated the dynamics of a two-level quantum emitter resonantly coupled to various protected flat bands, focusing on the role of disorder and on the nature of the lifted photonic mode that participates in the Rabi exchange.

We showed that weak hopping disorder produces a 
disorder-induced delocalization of the lifted mode. This effect reflects the competition between compact localization intrinsic to the flat band and Anderson localization \cite{li22,wang22,goda06, almeida23flat}.
Despite this crossover, we find that the atom–field coupling strength $\lambda$, and consequently the effective mode volume experienced by the emitter, is robust against the disorder. This stability arises because the local field amplitude at the emitter’s position $x_0$ 
is only weakly affected by the global reshaping of the lifted mode. Taken together with the large tolerance to emitter positioning inherent to flat bands \cite{wang25}, this makes flat-band cavities naturally resistant to fabrication imperfections.
%

%that distinguishes flat-band cavity QED from conventional photonic platforms, where disorder typically leads to large fluctuations in the coupling rate. 

%However, I'm doing it from the point of view of cavity QED dynamics. The state preparation, for instance, is based on a resonant jaynes-cummings-type interaction. And that is why I need chiral flat bands. I need to maintain the flat band even in the presence of disorder. Of course, this only works with off-diagonal disorder.

%%%%%%%%%%%%%%%%%%%%%
Our framework was illustrated for selected quasi-1D lattices comprising tens of cells, which are within the experimental capabilities of current photonic platforms \cite{vicencio15,mukherjee15,mukherjee15-2,real17,aravena22}. 
Note that in real photonics systems the interaction range is not necessarily restricted to nearest-neighbor coupling, rendering flat bands nor perfectly flat.
However, these unwanted long-range couplings can be suppressed
by, e.g., operating in regimes of weak inter-waveguide tunneling \cite{vicencio15,mukherjee15,hanafi22-2}.
Moreover, current photonic-lattice platforms offer local addressing and fine tunability \cite{danieli24}, which allow one to compensate for a small residual flat-band dispersion, including
scenarios where the on-site frequencies are not homogeneous.
We also remark that many highly-symmetric lattices are found to support flat bands
in the presence of next-nearest-neighbor (or even further) interactions \cite{morales16,mizoguchi19}.
Chiral flat bands also persist even in the presence of long-range interactions, as long as the chiral symmetry is preserved, meaning that no couplings occur within each sublattice.
In this case, however, 
CLSs may no longer be defined as the number of equations increases \cite{ramachandran17}. 

With respect to the resonant cavity-QED regime, we stress that the effective Rabi dynamics governed by Eq. (\ref{psi_lambda}) holds for any degenerate level that is separated from the rest of the spectrum by finite gap $\Delta\gg g_0$. If flat band acquires a small but finite bandwidth, the same framework continues to apply provided that the atom-field coupling $g_0$ is large compared to that bandwidth.
In future works, it should be interesting to investigate the consequences of a break in the chiral symmetry (by adding, e.g., diagonal disorder) in order to slightly lift the degeneracy of the flat band. 
%Small perturbations in the flat band yield regimes with ubiquitous transport properties \cite{} 

%%%%%%%%%%%%%%%%%%%%%%%%%
Cavity-QED offers a promising route for probing and preparing flat band states, particularly in scenarios where the lattice parameters are not well known. Note that the emitter locally couples to the photonic lattice without 
destroying its sublattice symmetry, regardless of the value of $g_0$. We enforce weak $g_0$ to neglect interaction with the dispersive modes and access
the localization characteristics of the flat band.  
In doing so, a single mode $\ket{\widetilde{\psi}}$ is lifted from the flat band due to the Rabi splitting but the zero-energy degeneracy is \emph{preserved}. Unless $U=1$, these remaining degenerate modes should be slightly modified around $x_0$. Indeed, the emitter stands as a lattice defect. 
 
From a broader perspective, the engineering of photonic reservoirs hosting flat bands \cite{leykam18-1,vicencio21} offers a range of opportunities across quantum technologies. Their unique transport properties enable robust long-range interactions between quantum emitters \cite{burillo20,dibenedetto25}, which can be harnessed for quantum communication protocols \cite{almeida23flat}. Moreover, photonic flat bands constitute a valuable resource for investigating open quantum system dynamics, including non-Markovian emission processes \cite{lorenzo17, monteiro25, monteiro25-2}.
These aspects will be subject of near-term investigations.

%Beyond the CLS description, characterize a flat band can be quite ambiguous due to the freedom in defining states within the degenerate subpace. Here, we have shown that the emission dynamics points to a preferable "direction". It actually becomes a $U=1$ CLS if the flat band allows for it. This adds upon them a fundamental trait which goes beyond the mere classifications of flat bands.  

\section*{Acknowledgements}

This work is supported by CNPq, CAPES and FAPEAL (Alagoas state agency).

%\textcolor{red}{NEW REFS \cite{nita13,maimaiti19,khomeriki16,wang22,goda06,ongoing,hanafi22-2,morales16,mizoguchi19,burillo20}}
%\textcolor{red}{Updated \cite{wang25}}

%\textcolor{green}{Added Figs. 1, 5, 6. Modified Fig. 3(a)}

%\textcolor{green}{We went from 4 pages to 7, excluding references, that is 3 new pages worth of content.}

%\bibliography{Refs}


\end{document}